# The Connection Between Political Stability and Inflation: Insights from Four South Asian Nations

Ummya Salma[1] and Md. Fazlul Huq Khan[2]

[1]Associate Professor, Department of Management Studies, Bangladesh University of Professionals (BUP), Email: ummya.salma@bup.org.bd

[2]PhD. candidate, University of International Business and Economics, Beijing
Email: Fazlul.khan@outlook.com

Abstract: This study explores the relationship between political stability and inflation in four South Asian countries, employing panel data spanning from 2001 to 2021. To analyze this relationship, the study utilizes the dynamic ordinary least square (DOLS) and fully modified ordinary least square (FMOLS) methods, which account for cross-sectional dependence and slope homogeneity in panel data analysis. The findings consistently reveal that increased political stability is associated with lower inflation, while reduced political stability is linked to higher inflation.

Keywords: Inflation, Political stability, DOLS, FMOLS.

## 1. Introduction

Political stability refers to the condition in which a country's government and political system are resilient and not subject to sudden or disruptive changes. It is a state of governance characterized by continuity, order, and the absence of significant political upheavals, such as coups, revolutions, or frequent changes in leadership. Hurwitz (1973)stated that the absence of violence and absence of structural change is a form of state stability.

Political stability is of paramount importance for both developing and developed nations, as the trajectory of economic growth is heavily reliant on a secure political environment. An unstable political climate breeds uncertainties that can dampen economic growth by discouraging investment. Moreover, subpar economic performance can result in government inefficiencies and political turmoil. (Alesina et al., 1996). Political stability often leads to the implementation of sound economic policies. When a country has stable and effective governance, it is more likely to adopt and maintain policies that control inflation.

Political stability can attract foreign investment and increase investor confidence. When investors believe that a country is politically stable, they are more willing to invest in that country's economy. This influx of foreign capital can stimulate economic growth and reduce inflationary pressures.

The problem of inflation, especially in developing nations, remains a topic of interest for both economists and non-economists. Inflation is the continuous rise in the overall price level, which diminishes the purchasing power of consumers within an economy (Tran, 2023). It leads to higher costs for goods and services and significantly affects various other economic factors. Inflation happens when the amount of money and credit in an economy increases more rapidly than the supply of goods and services. Most countries gauge inflation using the consumer price index (CPI) and the gross domestic product (GDP) deflator (Salma, 2021).

Although there is some disagreement regarding the specific mechanisms and pathways through which inflation manifests in the system, it is widely acknowledged that inflation is influenced by three fundamental sources: (a) excess aggregate demand (b) cost push factors and (c) Distributional factors (Durguti et al., 2021). When demand outpaces supply, it exerts upward pressure on prices. Similarly, increased market power of oligopolistic domestic firms, rising unit labor costs, higher prices for imported intermediate inputs, and occasional or systematic shortages of productive resources. Such factors contribute to inflation by increasing production costs. Prolonged conflicts can manifest in economic institutions and practices such as indexation and systematic price, wage, and rent adjustments, introducing unpredictable fluctuations into inflation trends. These factors collectively contribute to the dynamics of inflation within an economy.

Some researchers identified that maintaining consistently low and stable inflation has been associated with positive outcomes, including enhanced economic growth and development, greater financial stability, and a reduction in poverty (Ha et al., 2019) since persistent high inflation is consistently associated with subpar economic growth and the heightened risk of financial crises (Mishkin, 2008).

This study aims to make a valuable addition to the existing literature, as the earlier literature did not delve into the influence of political stability on inflation in the chosen countries. To identify the long-term relationship between the variables, we employ the dynamic ordinary least square (DOLS) and fully modified ordinary least square (FMOLS) estimation methods.

In addition to this introduction, the paper is structured into five sections. Section 2 provides a review of the literature. Section 3 covers the data and research methodology. Section 4 presents the empirical findings and discussions, while Section 5 offers conclusions and outlines policy implications.

## 2. Literature Review

A nation's economic performance is influenced not only by economic factors but also by political and institutional factors. Additionally, certain political and institutional variables are thought to have a significant role in elucidating both the variations and commonalities observed in economic outcomes across countries  Governments can potentially misuse monetary policy by pressuring monetary authorities to generate a monetary surprise to boost short-term output, leading to increased inflation without any tangible real benefits (Telatar et al., 2010).

There is a widely held belief that excessive inflation volatility can be detrimental to the overall economy (M. F. H. Khan, 2021). Nevertheless, despite this consensus, it is somewhat perplexing that numerous countries have grappled with both high and erratic inflation. Conversely, most developed nations have successfully maintained low and stable inflation rates.

Faruq (2023) identifies the factors that contribute to increased foreign direct investment (FDI) in a country and figured out that political stability attracts more foreign direct investment (FDI) which contributes to significantly lower inflation. Anjom & Faruq (2023) also reached a similar conclusion that financial and political stability is key in containing inflation to a satisfactory level.

Barugahara (2015) outlined three ways in which political instability can impact inflation volatility: (a) In countries marked by higher political instability, it is likely that tax evasion and the costs associated with tax collection will be more significant, (b) Political instability can also lead to reduced output and investment, resulting in a decrease in taxable assets and income. This can lead to an increased reliance on inflationary measures. Furthermore, political instability may contribute to larger fiscal deficits by reducing revenues and increasing public spending, particularly in countries with less developed financial markets. These deficits can have inflationary consequences (c) Political instability often hinders the effective implementation of coherent policies, eroding the government's competence and reducing its capacity to absorb shocks. This can ultimately lead to macroeconomic imbalances characterized by volatile inflation.

Researchers try to find the factors that reduced the inflation volatility. One of the factors is central bank independence which reduces inflation volatility (Cukierman et al., 1992). In developing countries that maintain fixed exchange rates, inflation levels and volatility tend to be lower (Bleaney & Fielding, 2002). Some researchers also find that trade openness reduces inflation volatility (Granato et al., 2006; Bowdler & Malik, 2017).

The majority of previous studies have aimed to establish a link between political instability and inflation, with only a limited focus on the relationship between political stability and inflation. Consequently, a positive correlation has been identified in studies examining the connection between political instability and inflation. However, researchers have uncovered a negative correlation when investigating the impact of political stability on inflation.

Political instability can influence inflation, and conversely, elevated inflation can also impact the political stability of a country. Paldam (1987) explores the impact of inflation on political change and the frequency of military regimes in eight Latin American countries from 1946 to 1984. The findings of this study suggest a bidirectional causal relationship. On one hand, a higher level of inflation corresponds to a greater frequency of military rule, even though military regimes are relatively unstable. On the other hand, political issues can contribute to higher inflation rates.

A similar outcome was discovered by Aisen & Veiga (2008) in their study spanning 160 countries from 1960 to 1999. They utilized a GMM estimator and observed that a higher degree of political instability is correlated with greater inflation rate volatility.

A decrease in political instability leads to a reduction in inflation. However, the presence of political freedom plays a significant role in shaping the connection between political instability and inflation. In a study conducted by Telatar et al. (2010), the impact of political instability and political freedom on inflation was explored. Their results indicate that reduced political instability is associated with lower inflation levels, but this effect is primarily observed in developed and low-inflation countries. However, when considering political freedom as a factor, political instability emerges as a significant influencer of inflation even in developing countries. Moreover, this influence is particularly pronounced in high-inflation countries.

Aisen & Veiga (2006) conducted a study examining the connection between political instability and inflation across 100 countries from 1960 to 1999, employing the system GMM estimator. The study's findings reveal that increased political instability is associated with higher inflation rates and greater seigniorage.

S. U. Khan & Saqib (2011) established a positive correlation between political instability and inflation through their analysis of the impact of political instability on inflation in Pakistan, utilizing data spanning from 1951 to 2007. Similarly, Haider et al. (2011) delved into the effects of political instability, governance, and bureaucratic corruption on inflation and growth in Pakistan. Their findings indicate that increased corruption coupled with weak governance leads to higher inflation and lower economic growth. They also note that poor governance and corruption coincide with political stability during democratic regimes.

Nonetheless, political instability can have contradictory effects on inflation. Specifically, instability in the political regime dimension tends to significantly elevate inflation, while instability in the government dimension significantly decreases inflation. (Ghanayem et al., 2023).

Political stability may exhibit a negative correlation with inflation when the shadow economy remains limited in size. In their 2017 study, Mazhar & Jafri (2017) examined whether the presence of a shadow economy could diminish the impact of political stability on inflation. They utilized data from 122 countries spanning the period from 1999 to 2007. The study's findings indicate that a negative correlation between political stability and inflation is observed only when the size of the shadow economy remains relatively small.

Similarly, Tran (2023) also examine the impact of political stability on inflation in the presence of shadow economy in Southeast Asian countries during the period from 2000 to 2017. They find that the higher level of the shadow economy contributes to the increasing inflation rate. However, the more political stability leads to the lower level of inflation. In addition, the impact of political stability on inflation rate really depends on the size of the shadow economy.

## 3. Research Methodology and Data

### 3.1 Model:

In this study, we investigate the impact of political stability on the inflation rate. We employ Dynamic Ordinary Least Square (DOLS) and Fully Modified Ordinary Least (FMOLS) estimators on a dataset comprising four South Asian countries for the period spanning from 2001 to 2021. The equation used for this study is:

$$INF_{it} = \beta_0 + \beta_1 PS_{it} + \beta_2 TR_{it} + \beta_3 EG_{it} + \beta_4 GNE_{it} + \beta_5 DC_{it} + \beta_6 BM_{it} + \varepsilon_{it}$$

In this study, the variable "i" corresponds to countries, and "t" signifies time. The dependent variable, denoted as "INF," represents the inflation rate, while the independent variable "PS" stands for political stability. Additionally, the control variables include "TR" for trade openness, "EG" representing the economic growth which is measured by log GDP per capita (LGDPP), "GNE" for Gross National Expenditure, "DC" signifying domestic credit, and "BM" denoting broad money.

### 3.2 Source of Data:

In this study, we investigate 4 South Asian countries including Bangladesh, India, Sri Lanka and Pakistan from 2001 to 2021. All the data are collected from World Development Indicators (WDI) of World Bank. Table-1 recapitulates the variables used in this study.

Table-1: Description of variables and data source

| Variables | Acronym | Description | Data source |
|---|---|---|---|
| Inflation | INF | Inflation, consumer prices (annual %) | WDI |
| Political stability | PS | Political stability index | WGI |
| Trade openness | TR | Trade (% of GDP) | WDI |
| Economic growth | LGDPP | Natural logarithm of GDP per capita growth (annual %) | WDI |

| Gross national expenditure | GNE | Gross national expenditure (% of GDP) | WDI |
|---|---|---|---|
| Domestic credit | DC | Domestic credit to private sector by banks (% of GDP) | WDI |
| Broad Money | BM | Broad money growth (annual %) | WDI |

In this study, the dependent variable under consideration is inflation, while the independent variable of interest is political stability. Additionally, we incorporate several control variables, including Trade openness (Chhabra & Alam, 2020), Economic growth (Paul et al., 1997), Gross national Expenditure (Kandil & Morsy, 2011), Domestic credit (Nwachukwu et al., 2014), and Broad money (Moser, 1995).

Table 2 displays the descriptive statistics for the dataset in this study. The highest recorded inflation rate is 22.564 percent, while the lowest is 2.135 percent, with an average of 7.247 percent. As for political stability, the mean value is -1.329, the standard deviation is 0.69, the minimum value is -2.81, and the maximum value is 0.09.

Table-2: Descriptive Statistics

| Variable | Obs | Mean | Std. Dev. | Min | Max |
|---|---|---|---|---|---|
| INF | 80 | 7.247 | 3.62 | 2.135 | 22.564 |
| PS | 80 | -1.329 | .69 | -2.81 | .09 |
| TO | 80 | 41.065 | 12.164 | 24.702 | 79.483 |
| LGDPP | 80 | 7.174 | .633 | 6.011 | 8.387 |
| GNE | 80 | 105.782 | 2.718 | 99.06 | 113.686 |
| DC | 80 | 34.898 | 11.615 | 14.579 | 54.572 |
| BM | 80 | 15.365 | 6.22 | 6.525 | 49.983 |

INF: Inflation; PS: Political stability; TO: Trade openness; LGDPP: Economic growth; GNE: Gross national Expenditure; DC: Domestic credit; BM: Broad money

## 4. Results

### 4.1 Cross- Sectional Dependence Test

Cross-sectional dependence is a common occurrence in panel estimation. Neglecting to account for cross-sectional dependencies in regression can result in a loss of efficiency and produce invalid test statistics in the estimation process. able 3 displays the test results, indicating that the null hypothesis of no cross-sectional dependence is accepted, as the probabilities from all the tests are greater than 5 percent. Nevertheless, for this study, we will utilize a panel unit root test.

Table-3: Cross-sectional dependence test

| Test | Statistic | Prob. |
|---|---|---|
| Breusch-Pagan LM | 5.777609 | 0.4486 |
| Pesaran scaled LM | -0.064199 | 0.9488 |
| Pesaran CD | 0.567777 | 0.5702 |

**4.2 Panel unit root test**:

We utilize the panel unit root test to assess stationarity and ascertain the integration order of the variables employed in this paper. Specifically, we apply the Levin et al. (2002), Fisher (1932), and Im et al. (2003) tests to determine whether the data exhibit stationarity at the level or after first differencing. The results in Table 4 indicate that all variables are stationary either at the level or after taking the first difference, as confirmed by all three tests.

Table-4: Panel unit root test

| Variables | Levin, Lin, and Chu | | Fisher | | IPS | |
|---|---|---|---|---|---|---|
| | I (0) | I (1) | I (0) | I (1) | I (0) | I (1) |
| INF | -1.32 | -6.23*** | 3.20*** | - | -2.01*** | - |
| PS | -0.53 | -1.82** | -0.84 | 15.58*** | 0.55 | -3.53*** |
| TR | -1.81** | - | -0.31 | 11.39*** | -0.37 | -3.93*** |
| LGDPP | -1.85** | - | 1.68** | - | -0.35 | -3.24*** |
| GNE | -1.82** | - | 1.61** | - | -1.60 | -4.85*** |
| DC | -3.02** | - | -0.17 | 7.91*** | 0.46 | -3.46*** |
| BM | -0.54 | -4.81*** | 3.64*** | - | -1.97** | - |

Notes: **, *** significant at 5% and 1% level, respectively.

INF: Inflation; PS: Political stability; TO: Trade openness; LGDPP: Economic growth; GNE: Gross national Expenditure; DC: Domestic credit; BM: Broad money

**4.3 Panel Cointegration Test**

To investigate the presence of a long-run equilibrium relationship among the variables, we employ various panel cointegration tests developed by Pedroni (1999, 2004), Westerlund (2005), and Kao (1999). The findings are presented in Table 5, revealing the existence of a long-run relationship between the variables. This is evident as the null hypothesis of no cointegration is rejected at a 5 percent significance level.

Table-5: Cointegration tests results

|  | Statistics |
|---|---|
| Pedroni |  |
| Modified Phillips-Perron t | 2.2447** |
| Phillips-Perron t | -5.4448*** |
| Augmented Dickey-Fuller t | -3.9693*** |
| Kao |  |
| Modified Dickey-Fuller t | -1.4971** |
| Dickey-Fuller t | -1.6844** |
| Augmented Dickey-Fuller t | -0.6223** |
| Unadjusted modified Dickey-Fuller t | -6.9983*** |
| Unadjusted Dickey-Fuller t | -4.7751*** |
| Westerlund |  |
| Variance ratio | -2.9797 ** |

Notes: *, **, *** significant at 10%, 5% and 1% level, respectively.

### 4.4 Empirical findings:

In this study, we aim to examine the impact of political stability on inflation in four South Asian countries. We employ Ordinary Least Squares (OLS), the Dynamic Ordinary Least Squares (DOLS) estimator recommended by Kao & Chiang (2000), and the Fully Modified Ordinary Least Squares (FMOLS) estimator technique proposed by Phillips & Hansen (1990) to assess the relationship between political stability and inflation in these selected nations. The results from all three tests consistently reveal a negative and statistically significant relationship between political stability and inflation in these countries. In other words, when political stability is higher, inflation tends to be lower, and vice versa.

Furthermore, our analysis indicates that an increase in economic growth and trade openness is associated with a reduction in the level of inflation. However, there is a positive and statistically significant relationship between gross national income and domestic credit provided to the private sector by banks and inflation. In a similar vein, while there is a positive relationship between broad money and inflation, it is not statistically significant.

Table-6: The impact of political stability on inflation

|  | OLS | DOLS | FMOLS |
| --- | --- | --- | --- |
| PS | -3.257*** | -3.050* | -3.983*** |
| TO | -0.066** | -0.145** | -0.003 |
| LGDPP | -0.382 | -4.172*** | -1.941* |
| GNE | 0.603*** | 0.631** | 0.517*** |
| DC | 0.114*** | 0.472*** | 0.257*** |
| BM | -0.030 | -0.102 | -0.028 |

Notes: *, **, *** significant at 10%, 5% and 1% level, respectively.
INF: Inflation; PS: Political stability; TO: Trade openness; LGDPP: Economic growth; GNE: Gross national Expenditure; DC: Domestic credit; BM: Broad money

## 5. Conclusion

This article presents an empirical analysis of the impact of political instability on inflation in Bangladesh, India, Pakistan, and Sri Lanka. Utilizing a balanced panel dataset spanning from 2001 to 2021 and employing the DOLS and FMOLS estimation techniques, this study reveals several key findings. Firstly, the results consistently demonstrate that political stability exerts a negative and significant influence on inflation in these selected countries. In other words, when political stability increases, inflation tends to decrease, and vice versa. Furthermore, the analysis highlights that trade openness and economic growth have a negative and significant impact on inflation. Conversely, gross national expenditure and domestic credit exhibit a positive and significant influence on inflation. While previous studies have explored the relationship between political stability and inflation to some extent, there is a notable scarcity of research that solely focuses on this effect in the selected countries. Therefore, this study contributes valuable insights to the field and can guide future research in this area.

The empirical evidence presented in this study underscores the importance of promoting political stability in the region. Policymakers should take into consideration the observed link between greater political stability and lower inflation when formulating policies. Additionally, fostering trade openness and stimulating economic growth while curbing national expenditure and domestic credit can be effective strategies for reducing inflation.